\documentstyle[preprint,aps,prb,tighten]{revtex}
\begin{document}

\widetext

\draft
\preprint{}
\title{Quantized Hall conductivity of Bloch electrons: \\
topology and the Dirac fermion\protect\thanks{To appear in Phys. Rev. B}
}
\author{Masaki Oshikawa\thanks{
Present Address: Department of Applied Physics, University of Tokyo.
Hongo, Bunkyo-ku, Tokyo 113 JAPAN.
}\thanks{E-mail address: {\tt oshikawa@mmm.t.u-tokyo.ac.jp}}
}

\address{
Institute for Solid State Physics, University of Tokyo \\
Roppongi, Minato-ku  Tokyo 106, Japan
}
\date{June 29, 1994}

\maketitle

\begin{abstract}
We consider the Hall conductivity of two-dimensional
non-interacting Bloch electrons when the magnetic flux per
unit cell is a rational number
$p/q$ where $p$ and $q$ are mutually coprime.
We present a counter-example for the naive expectation
that the Hall conductivity carried by a band is given by
treating gap minima as Dirac fermions.
Instead of the above expectation, we show that
the {\em change\/} of the Hall conductivity at a gap-closing
phenomenon is given by the Dirac fermion argument.
Comparing with the Diophantine equation, our result implies that
a band-gap closes at $q$ points simultaneously.
Furthermore,
we show that the dispersion relation is $q$-fold
degenerate in the magnetic Brillouin zone.
\end{abstract}


\narrowtext

\section{Introduction}
\label{sec:intro}

It was a surprise that Hall conductance of two-dimensional
electron system was found to be quantized
in some experiments.~\cite{vonKlitz}
The presence of disorder or edges are considered to be necessary
for the quantization of the Hall conductance~\cite{QHE} in the experiments.
However, basic important facts are that
there are energy gaps in the bulk system without disorder
and that the Hall conductivity is quantized when the Fermi level
lies in the gap.
In the free electron case,
it is easy to show this;
eigenstates are the degenerate Landau levels separated
by energy gaps, and each Landau level contributes $e^2/h$ exactly
to the Hall conductivity.

When a periodic potential is present, it is a more non-trivial problem
Since a Landau level
splits into several subbands by turning on a periodic potential,
one may expect that a subband carries fractional
(in units of $e^2/h$) Hall conductivity.
However, Thouless, Kohmoto, Nightingale and den Nijs~\cite{TKNN} (TKNN)
showed that
the Hall conductivity carried by a magnetic subband is always an integer.
This quantization comes from the topological nature;
TKNN integer is a topological
invariant on the magnetic Brillouin zone~\cite{KohTop}.
Ishikawa {\it et al.}~\cite{Ishikawaetal} also discussed
the topological aspect
of the quantized Hall conductivity.
Topological
character of the Hall conductivity
carried by edge states and its relation to the bulk TKNN integer
are also discussed~\cite{Hatsu:edge}.

On the other hand, it has been widely accepted that low-energy
behavior of a thermodynamic system is correctly described
by a continuum field theory.
In the present case, a minimum of a gap between magnetic bands may
be described by a Dirac fermion (in 2+1 dimensions), which gives
the Hall conductivity $- e^2/2h \mbox{ sgn}m $ ( $m$ is the Dirac mass.)
Hence we may expect that the Hall conductivity carried by a
band is given by counting only the Dirac fermions.
There are several works~\cite{Seme84,HaldQHE88,AnyonMott,Ishikawa84}
based on this idea.

However, it is not trivial whether the Dirac fermion argument is
true or not,
because the Hall conductivity is given by an integral over
whole Brillouin zone while the Dirac fermion argument looks only the
gap minima in the Brillouin zone.
Therefore, in this paper we investigate the validity
of the Dirac fermion argument for Bloch electrons,
and its relation to the topological TKNN integer.

Contrary to the naive expectation,
we found that the Hall conductivity carried by
a band cannot given by the Dirac fermion argument in general.
Instead, applying the argument by Simon~\cite{SimonQHE}
and others~\cite{Avronetal}, we showed that
the change of the Hall conductivity at a gap-closing point
is correctly described by the Dirac fermion argument.

Comparing with the Diophantine equation, this result implies
that a band-gap closes simultaneously at $q$ points.
Moreover,
we showed a stronger statement that the dispersion relation
is $q$-fold degenerate in the magnetic Brillouin zone
for general 2D Bloch electrons.
This is a generalization of the result~\cite{QHENN}
for the tight-binding
model on a square lattice.

Throughout this paper, we consider two-dimensional
non-interacting electron systems under a
periodic potential and a uniform magnetic field
at zero temperature.
The flux per unit cell is assumed to be a rational number,
and the Fermi level lies at a band gap except when the gap closes.
We set the velocity of light $c$ to be $1$ while we keep
Planck's constant $h$ or $\hbar=h/2\pi$.

The organization of this paper is as follows.
In Section~\ref{sec:Blochrev}, we give a brief review on
the Hall conductivity as a topological invariant.
In Section~\ref{sec:Diracrev}, we review the Dirac fermion in
$2+1$ dimensions.
Section~\ref{sec:failure} describes
a counterexample in which Dirac fermion argument
cannot give the correct value of the Hall conductivity of Bloch
electrons.
In Section~\ref{sec:change}, we show that the change of the Hall
conductivity during a gap-closing is given by the Dirac fermion argument.
A $q$-fold structure in the magnetic Brillouin zone is shown in
Section~\ref{sec:period}.
Conclusions and a discussion are given in Section~\ref{sec:concl}.

\section{Hall conductivity of Bloch electrons}
\label{sec:Blochrev}

Here we review known important facts about
the quantized Hall conductivity of Bloch electrons,
especially the topological aspect~\cite{TKNN,KohTop}
of the quantized Hall conductivity.

The Schr\"{o}dinger equation for a two-dimensional non-interacting
electron system in a uniform magnetic field is written as
\begin{equation}
H \psi(\vec{r}) =
 [ \frac{1}{2m}(\vec{p} +e \vec{A})^2 + U(\vec{r}) ] \psi(\vec{r}) =
 E \psi(\vec{r})
\end{equation}
where $\vec{p}$ is the momentum $ -i \hbar \nabla $ and
$\vec{A}$ is the vector potential.
In our case, the potential $ U(\vec{r}) $ is periodic, i.e.,
\begin{equation}
  U(\vec{r}+\vec{a}_1) = U(\vec{r}+\vec{a}_2) = U(\vec{r})
\end{equation}
where $\vec{a}_1$ and $\vec{a}_2$ are linearly independent
Braveis vectors.
We take the symmetric gauge $\vec{A} = 1/2 (\vec{B} \times \vec{r})$
for the moment.
Let us define the magnetic translation operators
\begin{equation}
\hat{T}_{\vec{R}} =
        \exp{ [ \frac{i}{\hbar} \vec{R} \cdot (\vec{p} - e\vec{A}) ]}.
\end{equation}
In the symmetric gauge, this operator commutes with the kinetic term
in the Hamiltonian, as well as with the potential term.

We consider the case where the magnetic flux per unit cell is
$p/q$, where $p$ and $q$ are mutually prime integers.
Namely, the magnetic flux density $\vec{B}$ satisfies
\begin{equation}
  \vec{B} \cdot (\vec{a}_1 \times \vec{a}_2) = \frac{p}{q} \frac{h}{e}.
\end{equation}
Then the following relations hold:
\begin{equation}
  [ \hat{T}_{q\vec{a}_1} , \hat{T}_{\vec{a}_2} ] = 0,
  [ \hat{T}_{\vec{R}} , H ] =0
\end{equation}
where $\vec{R} = n (q\vec{a}_1) + m \vec{a}_2$ and $n,m$ are integers.
Thus we can apply the Bloch's theorem if
we take an enlarged unit cell (magnetic unit cell)
which is $q$ times larger than the original unit cell.
Correspondingly, the reciprocal space (magnetic Brillouin zone)
becomes $1/q$ of the original Brillouin zone.
Namely, the Schr\"{o}dinger equation can be reduced as
\begin{equation}
  \hat{H}(\vec{k}) u^i_{\vec{k}}(\vec{r})
  = E^i u^i_{\vec{k}}(\vec{r})
\label{eq:redSch}
\end{equation}
\begin{equation}
  \hat{H}(\vec{k}) =
         \frac{1}{2m} ( - i \hbar \nabla + \hbar \vec{k} + e\vec{A} )^2
         + U(\vec{r})
\label{eq:Blochham}
\end{equation}
where $i$ is the band index,
$\vec{k}$ is the crystal momentum
and $u^i$ is a wavefunction with the generalized Bloch condition:
\begin{eqnarray}
u^i(\vec{r}) & =&
u^i(\vec{r}+ q \vec{a}_1)
\exp{[ - i \frac{qe}{2\hbar} \vec{B} \cdot ( \vec{r} \times \vec{a}_1)]}
\nonumber \\
&=&
u^i(\vec{r}+ \vec{a}_2)
\exp{[ - i \frac{e}{2\hbar} \vec{B} \cdot ( \vec{r} \times \vec{a}_2)]}.
\end{eqnarray}

Defining the reciprocal lattice vectors $\vec{g}_i$ by
$  \vec{g}_i \cdot \vec{a}_j = \delta_{ij}$,
we have equivalence relations
\begin{equation}
  \vec{k} \sim \vec{k} + \frac{2 \pi}{q} \vec{g}_1
          \sim \vec{k} + 2 \pi \vec{g}_2
\end{equation}
among crystal momenta $\vec{k}$.
Hence we can restrict $\vec{k}$ to the magnetic Brillouin zone
\begin{equation}
\label{eq:defMBZ}
\vec{k} = k_1 \vec{g}_1 + k_2 \vec{g}_2 \;\;\;\;\;
 ( - \frac{\pi}{q} \leq k_1 < \frac{\pi}{q},
   -\pi \leq k_2 < \pi ).
\end{equation}

{}From the linear response theory (Nakano-Kubo formula),
the Hall conductivity of the system is given by
\begin{eqnarray}
  \sigma_{xy} = && \sum_{i | E_i < E_F}
    \frac{e^2}{h} \frac{1}{2\pi i} \nonumber \\
&&    \times \int d^2k \int_{\mbox{\scriptsize MUC}}
     [ ( \frac{\partial u^i_{\vec{k}}(\vec{r})}{\partial k_y})^*
        \frac{\partial u^i_{\vec{k}}(\vec{r})}{\partial k_x}  -
      ( \frac{\partial u^i_{\vec{k}}(\vec{r})}{\partial k_x})^*
        \frac{\partial u^i_{\vec{k}}(\vec{r})}{\partial k_y}  ].
\end{eqnarray}
That is, the total Hall conductivity $\sigma_{xy}$ is obtained
as a sum of the contributions from all bands below the Fermi level.

The contribution from a single band can be written in a compact
form as follows.
We will omit the band index when a single band is considered.
For each band, we define a vector field
in the magnetic Brillouin zone by
\begin{equation}
  \hat{A}(\vec{k}) =
  \langle u(\vec{k}) | \nabla_k | u(\vec{k}) \rangle
  \equiv \int_{MUC} d^2 r u^*_{\vec{k}}(\vec{r})
                         \nabla_k u_{\vec{k}}(\vec{r})
\end{equation}
where $\nabla_k$ is a vector operator
$(\partial / {\partial k_x}, \partial / \partial k_y)$ and
MUC represents the magnetic unit cell.
It should be noted that we can choose the phase of the wavefunction
for each $\vec{k}$ arbitrarily.
Transformation of the phase as
\begin{equation}
  u'_{\vec{k}}(\vec{r}) =
       u_{\vec{k}}(\vec{r}) e^{i f(\vec{k})},
 \;\;\;\;
  \hat{A}'(\vec{k}) = \hat{A}(\vec{k}) + i \nabla_k f(\vec{k})
\end{equation}
is nothing but a $U(1)$ gauge transformation;
$\hat{A}(\vec{k})$ is a $U(1)$ gauge field
defined on the magnetic Brillouin zone.
Using this gauge field, the Hall conductivity carried by a single
band is written as
\begin{equation}
\label{eq:sigmaxy}
  \sigma_{xy} =
  \frac{e^2}{h} \frac{1}{2\pi i}
  \int_{MBZ} d^2 k \nabla_k \times \hat{A}(\vec{k})
\end{equation}
where $\nabla_k \times$ denotes the rotation in two-dimensional
$\vec{k}$-space and MBZ represents the magnetic Brillouin zone.

Although a naive application of the Stokes' theorem implies
that (\ref{eq:sigmaxy}) is zero,
a non-zero value of the hall conductivity
arises from a non-trivial topology of the fiber bundle.
That is,
one cannot determine the phase of the
wavefunction
$u_{\vec{k}}$ smoothly and uniquely over the entire Brillouin zone
in general.
For example,
one can fix the gauge by requiring that the amplitude
$ \langle a | u(\vec{k}) \rangle$
is real ($| a \rangle$ is an arbitrary wavefunction.)
Though this seems a well-defined gauge fixing, this cannot fix the
gauge at the zeros of the amplitude.
We must cover the region near a zero by another patch,
in which a different gauge fixing is chosen.
Phase mismatch between the two gauges produces a non-zero
value of (\ref{eq:sigmaxy}).
We call a zero of the amplitude as a vortex.
Uniqueness of the phase in each patch implies that the phase
mismatch around a vortex should be an integral multiple of $2\pi$.
This integer is called vorticity.
Thus the Hall conductivity carried by a band (\ref{eq:sigmaxy})
is $e^2/h$ times the total vorticity in the magnetic
Brillouin zone.
While the location of the vortices depends on the gauge,
total vorticity is gauge invariant and is
known as the first Chern number of the principal $U(1)$ bundle.

\section{Dirac fermion in 2+1 dimensions}
\label{sec:Diracrev}

To fix the convention,
here we briefly review the
relativistic Dirac fermion in $2+1$ dimensions.
The Lagrangian density is given by
\begin{equation}
\label{eq:DiracLag}
  {\cal L} =
   \bar{\psi} (i \hbar \partial_{\mu} - e A_{\mu} ) \gamma^\mu \psi
   - m \bar{\psi} \psi
\end{equation}
Where $A_{\mu}$ is the (background) vector potential.
The $\gamma$-matrices satisfy the anticommutation relation
$  \{ \gamma^{\mu} , \gamma^{\nu} \} = \eta^{\mu \nu}$
where $\eta^{\mu \nu}$ is the flat Minkowski metric.
In the 2+1 dimensions, $\gamma$-matrices are $2 \times 2$
matrices. We choose the convention
\begin{equation}
  \gamma^0 = \sigma^z , \gamma^1 = i \sigma^y,
  \gamma^2 = - i \sigma^x
\label{eq:gammaconv}
\end{equation}
where $\sigma^{x,y,z}$ are the standard Pauli matrices.
When the electromagnetic field is absent,
the 1-body Hamiltonian is derived from~(\ref{eq:DiracLag}) as
\begin{equation}
\label{eq:DiracHam2}
  H = m \sigma^z +  p_x \sigma^x + p_y \sigma^y
\end{equation}
where $(p_x, p_y)$ represents the momentum.

In the low-energy limit, the current is evaluated
by several methods~\cite{QHEDirac,AnyonMott,Ishikawa84}
\begin{equation}
  \langle e j^{\mu} \rangle =
    e \langle \bar{\psi} \gamma^{\mu} \psi \rangle =
    - \frac{e^2}{4h} \epsilon^{\mu \nu \sigma} F_{\nu \sigma}
    \mbox{sgn } m
\end{equation}
where $F_{\nu \sigma}$ is the field strength
$ \partial_{\nu} A_{\sigma} - \partial_{\sigma} A_{\nu}$.
Namely, the Hall conductivity carried by a Dirac fermion is
$- e^2/2h \mbox{sgn }m$.

\section{Failure of the Dirac fermion argument}
\label{sec:failure}

In many cases, the low-energy behavior of a thermodynamic system
can be treated by a continuous field theory.
{}From the field-theory point of view, this means a
field theory is independent of details in the regularization.

In the case of the Bloch electron system in a magnetic field,
the low-energy states near a gap minimum
may be treated as a Dirac fermion~\cite{NoteDirac}.
Thus we may expect that summing up contributions from each
Dirac fermion gives the Hall conductivity of a band.
Several authors made discussions based on this idea, and
in fact it seems correct
in some simple lattice models~\cite{Seme84,HaldQHE88}.

However, we show that this naive expectation is false in
general.
As a counterexample, we take a tight-binding model discussed in
Ref.~\onlinecite{QHENNN}.
We consider a isotropic square lattice with nearest-neighbor
(NN) and next-nearest-neighbor (NNN) hoppings with
$1/3$ magnetic flux per unit cell.
Let the absolute value of the NN and NNN hoppings be 1 and $t_c$,
respectively.
Namely, the Hamiltonian of the system reads

\begin{eqnarray*}
  H = &&\sum_{n,m} c^{\dagger}_{n+1,m} c_{n,m}
           \exp{(i \theta_{A})}
      + \sum_{n,m} c^{\dagger}_{n,m+1} c_{n,m}
           \exp{(i \theta_{B})} \\
     && + t_c \sum_{n,m} c^{\dagger}_{n+1,m+1} c_{n,m}
           \exp{(i \theta_{C})} \\
     && + t_c \sum_{n,m} c^{\dagger}_{n,m+1} c_{n+1,m}
           \exp{(i \theta_{D})}
     + \mbox{H.c.}
\end{eqnarray*}
where
\begin{equation}
  \theta_{A} = 0, \;\; \theta_{B} = \frac{2 \pi}{3} n, \;\;
  \theta_{C} = \theta_{D}
          = \frac{2 \pi}{3}(n+\frac{1}{2}).
\end{equation}
The energy spectrum of this model is analyzed by Hatsugai and
Kohmoto~\cite{QHENNN}.
According to them, the energy eigenvalue $E$ for each (crystal) momentum
$\vec{k}$ is determined by the equation
\begin{equation}
  F(E) = f(\vec{k})
\end{equation}
where $F(E)$ is some polynomial of $E$ and $f(\vec{k})$ is given by
\begin{eqnarray}
  f(\vec{k}) = && 2 \cos{(3k_x)}[ 1 - 3 {t_c}^2 ]
                  + 2 \cos{(3 k_y)} [1 - 3 {t_c}^2 ] \nonumber \\
          &&  -2 {t_c}^3 \{ \cos{[3(k_x+k_y)]} + \cos{[3(k_x-k_y)]} \} .
\end{eqnarray}

Precisely speaking, a minimum of a energy gap cannot always be regarded
exactly as a Dirac fermion.
In general Bloch electron systems, there are several bands separated
by the energy gaps.
If the gap is
very small compared with other gaps, we can ignore other bands.
Then the low-energy states near a gap minimum can be considered
as one-particle states of a Dirac fermion.
It is possible that a gap-closing point become singular; in this case
the dispersion of the low-energy states is not relativistic.
We have checked this is not the case in our counter-example.
We will make a detailed
discussion on this point in the next section.

In our model, the first gap closes simultaneously at three points
when $t_c \sim 0.268$.
An extremal point of an energy band is given by an
extremal point of $f(\vec{k})$, and
it is easy to show that the first gap has three minima
in the neighborhood of the gap-closing point $t_c \sim 0.268$.
Thus there are just three Dirac fermions
in this region, if the Fermi level lies in the first gap.
The dispersion relation of the model $t_c = 0.25$, which is
near the gap-closing point, is shown in Fig.~1.

Since a Dirac fermion contributes $ \pm e^2/2h$
to the Hall conductivity $\sigma_{xy}$,
it should be half-odd-integral multiple of $e^2/h$
from the Dirac fermion argument.
However it should be an integral multiple of $e^2/h$
according to the general theory of TKNN.
Hence the naive Dirac fermion argument is incorrect.
In other words, there is no
`anomaly-cancelling partner'~\cite{HaldQHE88} in our model.

\section{Change of the Hall conductivity and the Dirac fermion}
\label{sec:change}

The naive expectation that the Dirac fermion argument
gives the Hall conductivity carried by a
band is disproved in the last section.
Nevertheless, the Dirac fermion argument still makes sense
for our problem.
We proved the following proposition.

\begin{quote}
When a parameter in the Hamiltonian is varied,
the Hall conductivity changes only where the gap closes.
This {\em change\/} of the Hall conductivity is
correctly given by
the Dirac fermion argument;
the Hall conductivity decreases by $e^2/h$ if the Dirac
mass changes from negative to positive, and increases
by $e^2/h$ if it changes from positive to negative.
If the gap closes at several points simultaneously,
the change of the Hall conductivity is given by
summing up the contributions from each gap-closing point.
\end{quote}

The above proposition
is shown by mapping of the low-energy states to
the Dirac fermion and
calculation of the change of the Chern number.
The important point is that the change of the Chern number is
determined only by the neighborhood of the gap-closing point.
This fact was noticed by Simon~\cite{SimonQHE},
and also by Avron {\em et al.\/}~\cite{Avronetal}
in the context of the network problem.
We also note that the proof has similar structure
to the ``intuitive topological proof''~\cite{Absence2}
of the Nielsen-Ninomiya theorem in $3+1$ dimensions.

To discuss the {\em change\/} of the Hall conductivity,
the Hamiltonian is assumed to vary smoothly.
Namely, we discuss a family of
Hamiltonians labelled by a parameter $\lambda$.
The reduced Hamiltonian $\hat{H}$ also should be parameterized
by $\lambda$
in the reduced Schr\"{o}dinger equation~(\ref{eq:redSch}).
Thus $\hat{H}$ depends on three parameters,
$k_x, k_y$ and $\lambda$;
it can be said that $\hat{H}$ is defined in a three-dimensional
space whose coordinates are $k_x, k_y$ and $\lambda$.

First let us focus on the gap-closing phenomena.
Assume that there is a gap-closing point $(\vec{k}^*,\lambda^*)$
where energies of two bands are degenerate.
Here we consider the generic case in which only two bands touch
at this point.
We can neglect other bands near this gap-closing point.
Furthermore, we expand the Hamiltonian to the first order in
\begin{equation}
  {\bf p} = ( p_x , p_y , p_z)
                = ( \hbar (k_x - {k_x}^*), \hbar (k_y - {k_y}^*),
                 \lambda - \lambda^* )
\end{equation}
where ${\bf  p}$ is a three-dimensional vector.
Since the effective Hamiltonian for the two bands is $2 \times 2$ Hermitian
matrix for each $\bf p$, the most general form of the expansion is
\begin{equation}
\label{eq:effHam2}
 \hat{H} = E^*  + {\bf b} \cdot {\bf  p}
          +  V_{\mu}^{\nu} \sigma^{\mu} p_{\nu}
          + O({\bf  p}^2)
\end{equation}
where $\mu, \nu = 1,2,3$ and $\sigma^{1,2,3}=\sigma^{x,y,z}$.
If $V$ is singular ($\det{V} =0$), the dispersion relation near
the gap-closing point is not Dirac-fermion like.
We don't consider such non-generic~\cite{Avronetal} cases in this paper.

In order to see this is a Hamiltonian of the Dirac fermion,
we make a unitary transformation and a redefinition of
$(p_x, p_y)$.
Since
the term proportional to the identity matrix can be absorbed
to the redefinition of the energy and hence irrelevant,
we only consider the terms containing a Pauli matrix.
It is noted that
a $SU(2)$ transformation $U$ induces the transformation on
the Pauli matrices as
\begin{equation}
\label{eq:su2adj}
  U \sigma^{\mu} U^{-1} = R^{\mu}_{\nu} \sigma^{\nu}
\end{equation}
where $R^{\mu}_{\nu}$ is a $SO(3)$ matrix.
There is always an appropriate $SU(2)$ matrix $U$ corresponds to
some given $R$.
Any regular matrix $V$ can be decomposed as
(Gram-Schmidt decomposition)
\begin{equation}
\label{eq:GramSchm}
  V = O T
\end{equation}
where $O$ is an orthogonal matrix and $T$ is an upper triangle
matrix with positive diagonal elements.
Let us choose $R^{-1} = O W$ where
$W = \mbox{diag} ( 1, 1, \mbox{sgn}(\det{V}) )$.
Then $U$ transforms the Hamiltonian as
\begin{equation}
  U \hat{H} U^{-1} \sim
     \sigma^{\mu} T_{\mu}^{\rho} W_{\rho}^{\nu} p_{\nu}
\end{equation}
where the terms proportional to the identity matrix are omitted.
Let us define
\begin{equation}
  \tilde{p}_{\mu} = (  \tilde{p}_x, \tilde{p}_y,
                      \tilde{p}_z ) =
  T_{\mu}^{\nu} p_{\nu}.
\end{equation}
Since $T$ is an upper triangle matrix with positive
diagonal elements,
this is a parity-conserving Affine transformation
on $p_x, p_y$ and a scale transformation on
$p_z$: $\tilde{p}_z = (\mbox{positive constant}) \times p_z$.

By this redefinition, the Hamiltonian becomes
\begin{equation}
\label{eq:effHam2a}
  \hat{H} \sim \mbox{sgn}(\det{V}) \tilde{p}_z \sigma^z
             + \tilde{p}_x \sigma^x
             + \tilde{p}_y \sigma^y,
\end{equation}
which is nothing but the Dirac Hamiltonian~(\ref{eq:DiracHam2}) with
the mass $\mbox{sgn}(\det{V}) \tilde{p}_z$.

In this way, the neighborhood of a gap-closing point can be
regarded as a Dirac fermion and the sign of the mass is
opposite before and after the gap-closing.
In Ref.~\onlinecite{Absence2}, the Hamiltonian which is formally same
as~(\ref{eq:effHam2a}) represents the one-body Weyl Hamiltonian
in 3+1 dimensions.
In this case, $\tilde{p}_z$ denotes the third momentum,
and the Weyl fermion is left-handed (right-handed) if $\det{V}>0$
($\det{V}<0$).
The similarity between our argument and that of Ref.~\onlinecite{Absence2}
is based on the formal similarity between
the one-body Dirac Hamiltonian in 2+1 dimensions
and the one-body Weyl Hamiltonian in 3+1 dimensions.
However, for our proof
we should be careful not to mix the parameter $\lambda$
with the momenta $k_x, k_y$ during the transformation.
This point has been solved by the Gram-Schmidt
decomposition.

However, in our problem,
we should be careful not to mix the parameter $\lambda$
with the momenta $k_x, k_y$ during the transformation
because each section for constant $\lambda$ represents
a physical model.
This point has been cleared by use of the Gram-Schmidt
decomposition~(\ref{eq:GramSchm}).

Next, we calculate the change of the Chern number as
discussed in Refs.~\onlinecite{SimonQHE,Avronetal}.
We introduce vortex lines in the three-dimensional parameter space.
The vortex lines are defined by
\begin{equation}
\label{eq:curvedef}
  \{ (k_x, k_y, \lambda) | \exists j :
       \langle a |  u^j(\vec{k};\lambda) \rangle =0 \}
\end{equation}
with respect to some reference vector
(in the reduced Hilbert space) $| a \rangle$.
The real part and the imaginary part of
the above condition fix two degrees of freedom.
Thus (\ref{eq:curvedef}) defines a set of curves (vortex lines)
in the three-dimensional parameter space.
When a vortex line satisfies $\langle a | u^i(\vec{k};\lambda) \rangle =0$
for the band $i$,
the vortex line is said to be in the $i$-th band.

It should be noted that
the definition of the vortex lines is similar to that of the vortices.
In fact, a section of the three-dimensional parameter space
for a certain constant $\lambda$ represents the magnetic Brillouin zone,
and its intersection with the vortex lines are vortices.
We define the orientation of a vortex line as shown in Fig.~2.
it is taken in such a way that $\lambda$ increases (decreases)
if the vorticity at the section is positive (negative).

The following propositions~\cite{Absence2}
are central for our result.
\begin{itemize}
\item For each gap-closing point, there is always a vortex line
passes through it.
\item The vortex line passes through
the gap-closing point upward (i.e.\ from the lower band to
the upper band) along the defined orientation
if $\det{V} > 0$ and downward if $\det{V} <0$.
\end{itemize}

Now we are going to examine the change of the Hall conductivity
at the gap-closing point.
We assume that the Fermi energy always lies in the gap
except when it closes.
We consider the case $\det{V} >0$.
Let us consider the vortex line which passes through the
gap-closing point.
If the orientation of the vortex line at the
gap-closing point is in the direction that $\lambda$ increases,
the local vorticity is $+1$.
According to the above proposition,
the vortex line lies in the lower band when $\lambda < \lambda^*$
and in the upper when $\lambda > \lambda^*$.
(See Fig.~3.)

If the orientation of the vortex line at the gap-closing
point is the reverse, i.e.\ in the direction that $\lambda$ decreases,
the vorticity is $-1$.
Thus in this case
the vortex line lies in the upper band when $\lambda < \lambda^*$
and in the lower when $\lambda > \lambda^*$.
When $\lambda$ is increased through the gap-closing value $\lambda^*$,
the total vorticity of the lower band decreases by $1$ in the cases;
the Hall conductivity decreases by $e^2/h$
irrespective of the direction of the vortex line.

On the other hand,
the mass of the Dirac fermion is negative when $\lambda < \lambda^*$.
If $\lambda$ is increased,
the mass become positive when $\lambda > \lambda^*$.
The prediction of the Dirac fermion argument is that
the Hall conductivity decreases by $e^2/h$
when $\lambda$ is increased through the gap-closing value $\lambda^*$.
Now we can see it gives the correct change of the Hall conductivity.
Similar arguments can be applied for a gap-closing point
with $\det{V} <0$.
It is also easy to see that
the change of the Hall conductivity is given by
summing up the contributions from each gap-closing point
when the two bands touch simultaneously at several points.

We also note that, the vortex only moves to another band and
never appear or disappear when two bands touch.
This implies that the sum of the Hall conductivity carried
by two bands is conserved in a gap-closing.
This is a simple illustration for the conservation law
first discussed by Avron, Seiler and Simon~\cite{Avron83}.

Let us show how our result applies to
the example presented in Section~\ref{sec:failure}.
The first gap closes simultaneously at three points
when $t_c = t_c^* \sim 0.268$.
The matrix $V$ for these points can be calculated numerically
and we obtain $\det{V} \sim 7.2 >0$ for each point
(see also the next section).
Thus the Dirac mass is negative when $t_c <t_c^*$ and positive
when $t_c > t_c^*$, and the Hall conductivity of the fist band
decreases by $3e^2/h$ when $t_c$ is increased through $t_c^*$.
This is consistent with the result
$\sigma_{xy}$ changes from $e^2/h$ to $-2e^2/h$
obtained by Hatsugai and Kohmoto~\cite{QHENNN}.

\section{Periodic structure in the magnetic Brillouin zone}
\label{sec:period}

It is known
that the Hall conductivity of the Bloch electrons
satisfies the Diophantine equation~\cite{TKNN,Diophantine}
\begin{equation}
\label{eq:Diophantine}
  r = q s_r + p t_r
\end{equation}
where the Fermi level lies at the $r$-th gap, $s_r$ is an integer and
the total Hall conductivity is $t_r e^2/h$.
A gap-closing does not affect the value of $p,q$ and $r$.
Thus if $t'_r e^2/h$ is the Hall conductivity after the gap-closing,
we have
\begin{equation}
  q ( s_r - s'_r ) + p (t_r - t'_r ) =0
\end{equation}
for an integer $s'_r$;
the change of the Hall conductivity
should be an integral multiple of $ q e^2/h$.

Comparing with this, our result implies that two bands touch
simultaneously at multiple of $q$ points.
In fact, in our example shown in Section~\ref{sec:failure}
the first gap closes simultaneously at $q=3$ points.
For the tight-binding model on a square lattice,
Kohmoto~\cite{QHENN} showed that
the dispersion relation is $q$-fold in the magnetic Brillouin zone,
combining a duality transformation~\cite{duality} and a gauge transformation.
Although it seems difficult to extend
the duality transformation to general models,
we show that the $q$-fold structure
is common in general Bloch electron systems.

For convenience, we make a gauge transformation in the reduced
Schr\"{o}dinger equation~(\ref{eq:redSch}) to a Landau-like gauge as
\begin{eqnarray}
  u'_{\vec{k}}(\vec{r})
 &=& \exp{[ -i\pi\frac{p}{q} (\vec{r} \cdot \vec{g}_1)
          (\vec{r} \cdot \vec{g_2})]} u_{\vec{k}}(\vec{r}) \\
\vec{A}' & =& \frac{1}{2} (\vec{B} \times \vec{r})
 - \frac{1}{2}[\vec{r} \cdot (\vec{B} \times \vec{a}_2) \vec{g}_2
             - \vec{r} \cdot (\vec{B} \times \vec{a}_1) \vec{g}_1]
\nonumber \\
&=& [(\vec{B} \times \vec{r}) \cdot \vec{a}_2] \vec{g}_2.
\end{eqnarray}

If we shift the momentum by
$\vec{k} \rightarrow \vec{k} + 2\pi (p/q) \vec{g}_2$,
the kinetic term changes as
\begin{equation}
  \frac{1}{2m}( - i\hbar \nabla + \hbar \vec{k} + e \vec{A} )^2
\rightarrow
  \frac{1}{2m}( - i\hbar \nabla +
         \hbar \vec{k} + h \frac{p}{q}  \vec{g}_2 + e \vec{A} )^2.
\end{equation}
This can be absorbed by the shift
$\vec{r} \rightarrow \vec{r} - \vec{a}_1$, which changes the
vector potential as
\begin{eqnarray}
 \vec{A} & \rightarrow &
 \vec{A} -
  [(\vec{a}_1 \times \vec{a}_2)\cdot \vec{B}] \vec{g}_2  \nonumber \\
& = & \vec{A} - \frac{h}{e} \frac{p}{q} \vec{g}_2
\end{eqnarray}
This shift does not change the potential term.
Thus we see that
$E{\bf (}\vec{k} + 2 \pi (p/q) \vec{g}_2{\bf )} = E(\vec{k})$
Since $p$ and $q$ are coprimes, for an arbitrary integer $m$
there is always an integer $n$ so that
\begin{equation}
  n p \equiv m \;\;\; ( \mbox{mod}\;q)  .
\end{equation}
This leads to the symmetry
\begin{equation}
       E(\vec{k} + \frac{2\pi m}{q} \vec{g}_2 ) = E(\vec{k})  .
\end{equation}
Hence the above symmetry means that the dispersion relation
is $q$-fold;
the magnetic Brillouin zone~(\ref{eq:defMBZ})
consists of $q$ sub-zones which have the same dispersion relation.

The fact that two bands should touch simultaneously at multiple
of $q$ points follows from this proposition.
Moreover, we can also see that
the Dirac masses (if they are well-defined) are
the same for the $q$ Dirac fermions connected by the shift
$\vec{k} \rightarrow \vec{k} + 2 \pi (m/q) \vec{g}_2$.

The above result shows the consistency between the Dirac fermion argument
and the Diophantine equation.
We note that, in general, the Diophantine equation~(\ref{eq:Diophantine})
implies that the Hall conductivity is not an integral multiple
of $qe^2/h$ nor $qe^2/(2h)$,
although the dispersion is $q$-fold.
Hence a Dirac fermion argument cannot give the correct Hall
conductivity as long as one defines Dirac fermions with respect
to the dispersion relation.
The $q$-fold structure restricts only the {\em change\/} of the
Hall conductivity to be a multiple of $q e^2/h$.
While the Dirac fermion argument can determine the precise value of the
change of the Hall conductivity, the Diophantine equation
restricts also the value of the total Hall conductivity;
they are consistent and complementary results.

\section{Conclusions and Discussion}
\label{sec:concl}

We discussed two-dimensional non-interacting Bloch electrons
in a uniform magnetic field.
The naive expectation that the Hall conductivity
carried by a band is given by treating gap minima as Dirac fermions
was found to be false.
Instead of the naive expectation, we showed that the
change of the Hall conductivity when the band gap closes
is correctly given by the Dirac fermion argument.
Comparing with the Diophantine equation, our
result implies that the gap-closing occurs simultaneously
at multiple of $q$ points.
We proved a stronger statement that the magnetic Brillouin zone
consists of $q$ sub-zones that have the same dispersion relation.

Although there is no reason that a naive field-theory prediction should
be always true, one may ask why it fails in the present case.
Our answer is that the failure is already implicit in the (naive)
Dirac fermion
argument itself. The Hall conductivity of a Dirac fermion
depends only on the {\em sign\/} of the Dirac mass.
That is, a Dirac fermion with arbitrary large mass contributes
to the Hall conductivity.
However, the large Dirac mass corresponds to the large gap between
the bands in the Bloch electrons.
When the gap is not small compared with other gaps,
the identification between the Dirac fermion and the gap minimum
becomes ambiguous.

On the other hand,
since the Hall conductivity is given by an integral
over the whole magnetic Brillouin zone, it seems impossible
to determine the value only from gap minima.
However, the Hall conductivity is not an ordinary integral
but a topological invariant
and cannot change except when the gap closes.
Hence we can expect that the change of the Hall conductivity
can be described by the local neighborhood of the gap-closing point.
This expectation is realized in the proof.

Finally, we comment on the Dirac fermion argument~\cite{HaldQHE88} using the
Widom-Streda formula.
It is summarized as follows.
The Hall conductivity is related to the change of the charge density
in an infinitesimal extra magnetic field.
In a constant magnetic field,
the energy levels of a Dirac fermion
form Landau levels. There are zero modes, whose energy depends on the
Dirac mass, among the Landau levels.
Since the spectrum besides the zero modes is symmetric for the Dirac fermion,
the Hall conductivity is determined by the zero modes as
$ - e^2/(2h) \mbox{sgn}m$ for each Dirac fermion.
However, in a general Bloch electron system, the spectrum is {\em not\/}
symmetric about the Fermi level; the Dirac fermion argument is not
reliable to obtain the Hall conductivity.
Nevertheless,  we can expect that only the contribution from the
zero modes will change at a gap-closing point.
This is consistent with our result that the Dirac fermion argument
gives only the correct {\em change} of the Hall conductivity.

\acknowledgments

I would like to express my gratitude to Prof. Mahito Kohmoto for
stimulating discussions and encouragement.
It is also a pleasure to thank Prof. Yong-Shi Wu,
Prof. Kenzo Ishikawa, Dr. Yasuhiro Hatsugai,
Masahito Takahashi and Jun Nishimura for useful discussions.
This work was supported in part by Grant-in-Aid for Scientific Research
from the Japanese Ministry of Education, Science and Culture.
The support of the JSPS fellowships is gratefully acknowledged.


\begin{thebibliography}{99}

\bibitem{vonKlitz}
K. von Klitzing, G. Dorda and M. Pepper, Phys. Rev. Lett.
{\bf 45}, 494 (1980).

\bibitem{QHE}
for a review, see for example
R. E. Prange and S. M. Girvin (eds), {\em The Quantum Hall Effect\/},
Springer-Verlag (1988).

\bibitem{TKNN}
D. J. Thouless, M. Kohmoto, M. P. Nightingale, and M. den Nijs,
Phys. Rev. Lett. {\bf 49}, 405 (1982).

\bibitem{KohTop}
M. Kohmoto, Ann. Phys. (N.Y.) {\bf 160}, 343 (1985).

\bibitem{Ishikawaetal}
K. Ishikawa, Phys. Rev. Lett. {\bf 53}, 1615. \\
N. Imai, K. Ishikawa, T. Matsuyama and I. Tanaka, Phys. Rev. B
{\bf 42}, 10610. \\
K. Ishikawa, Prog. Theor. Phys. Suppl. {\bf 107},167.

\bibitem{Hatsu:edge}
Y. Hatsugai,
Phys. Rev. B {\bf 48}, 11851 (1993). \\
Y. Hatsugai, ``The Chern number and the Edge States in the Integer Quantum Hall
Effect'', MIT preprint.

\bibitem{Seme84}
G. W. Semenoff, Phys. Rev. Lett. {\bf 53}, 2449 (1984).

\bibitem{HaldQHE88}
F. D. M. Haldane, Phys. Rev. Lett. {\bf 61}, 2015 (1988).

\bibitem{AnyonMott}
W. Chen, M. P. A. Fisher and Y. S. Wu,
``Mott Transitions in An Anyon Gas'',
Univ. of British Columbia preprint UBCTP-92-28
(cond-mat/9301036).

\bibitem{Ishikawa84}
K. Ishikawa, Phys. Rev. D {\bf 31}, 1432 (1985).

\bibitem{SimonQHE}
B. Simon, Phys. Rev. Lett. {\bf 51}, 2167 (1983).

\bibitem{Avronetal}
J. E. Avron, A. Raveh and B. Zur, Rev. Mod. Phys. {\bf 60}, 873 (1988). \\
J. E. Avron and L. Sadun, Phys. Rev. Lett. {\bf 62}, 3082 (1989). \\
J. E. Avron and L. Sadun, Ann. Phys. (N.Y.) {\bf 206}, 440 (1991).


\bibitem{QHENN}
M. Kohmoto, Phys. Rev. B {\bf 39}, 11943 (1989).

\bibitem{QHEDirac}
A. Redlich, Phys. Rev. Lett. {\bf 52}, 18 (1984). \\
A. J. Niemi and G. W. Semenoff, Phys. Rev. Lett. {\bf 51}, 2077 (1983). \\
S. Deser, R. Jackiw and S. Templeton,
Phys. Rev. Lett. {\bf 48}, 975 (1982).

\bibitem{NoteDirac}
In the Dirac fermion argument, the uniform magnetic field in the
original electron model should not be taken as the background of the
Dirac fermion.
Its effect is already included in the Dirac fermion itself.

\bibitem{QHENNN}
Y. Hatsugai and M. Kohmoto, Phys. Rev. B {\bf 42}, 8282 (1990). \\
We note that the Hamiltonian in the present paper
is different by the overall factor $-1$
from that in the above reference.

\bibitem{Absence2}
H. B. Nielsen and M. Ninomiya, Nucl. Phys. B {\bf 193},173 (1981).

\bibitem{Avron83}
J. E. Avron, R. Seiler and B. Simon, Phys. Rev. Lett. {\bf 51}, 51
(1983).

\bibitem{Diophantine}
I. Dana, Y. Avron and J. Zak, J. Phys. C {\bf 18}, L679 (1985). \\
M. Kohmoto, J. Phys. Soc. Jpn. {\bf 61}, 2645 (1992).

\bibitem{duality}
S. Aubry and G. Andr\'{e}, Ann. Isr. Phys. Soc. {\bf 3},133 (1980).



\end{thebibliography}
\end{document}